\documentclass[12pt,letterpaper]{article}

\usepackage{natbib,epsfig,graphicx,setspace}
\usepackage{amsmath,amsthm,amssymb,enumerate,color}
\usepackage{mathrsfs}
\usepackage{bm,verbatim,subfigure}

\newcommand{\diag}{\text{diag}}
\newcommand{\e}{\text{E}}

\def\bbeta{{\mbox{\boldmath $\beta$}}}
\def\bgamma{{\mbox{\boldmath $\gamma$}}}
\def\bvarepsilon{{\mbox{\boldmath $\varepsilon$}}}

\def\bx{{\mbox{\boldmath $x$}}}
\def\bX{{\mbox{\boldmath $X$}}}
\def\bW{{\mbox{\boldmath $W$}}}
\def\by{{\mbox{\boldmath $y$}}}
\def\bz{{\mbox{\boldmath $z$}}}

\newcommand{\jasa}{\emph{Journal of the American Statistical Association}}

\numberwithin{equation}{section} 

\begin{document}

\title{Robust Linear Regression: A Review and Comparison}
\author{Chun Yu$^{1}$, Weixin Yao$^{1}$, and Xue Bai$^{1}$ \\
\\$^{1}$Department of Statistics, \\Kansas State University, Manhattan, Kansas, USA 66506-0802.\\
}
\date{}
\maketitle{}

\bigskip

\begin{abstract}
Ordinary least-squares (OLS) estimators for a linear model are very
sensitive to unusual values in the design space or outliers among
$y$ values. Even one single atypical value may have a large effect
on the parameter estimates. This article aims to review and describe
some available and popular robust techniques, including some recent
developed ones, and compare them in terms of breakdown point and
efficiency. In addition, we also use a simulation study and a real
data application to compare the performance of existing robust
methods under different scenarios.



\end{abstract}

\vskip 20pt


\section{Introduction}
Linear regression has been one of the most important statistical
data analysis tools. Given the independent and identically
distributed (iid) observations $(\bx_{i},y_{i})$, $i = 1,\ldots,n$,
in order to understand how the response $y_i$s are related to the
covariates $\bx_i$s,  we traditionally assume the following linear
regression model
\begin{equation}
y_{i}=\bx_{i}^T\bbeta+\varepsilon_{i},
\end{equation}
where $\bbeta$ is an unknown $p\times 1$ vector, and the
$\varepsilon_{i}$s are i.i.d. and independent of $\bx_{i}$ with
$\e(\varepsilon_{i}\mid\bx_i)=0$. The most commonly used estimate
for $\bbeta$ is the ordinary least square (OLS) estimate which
minimizes the sum of squared residuals
\begin{equation}
\sum_{i=1}^{n}(y_{i}-\bx_{i}^T\bbeta)^{2}. \label{lse}
\end{equation}
However, it is well known that the OLS estimate is extremely
sensitive to the outliers. A single outlier can have large effect on
the OLS estimate.

In this paper, we review and describe some available robust methods.
In addition, a simulation study and a real data application are used
to compare different existing robust methods. The efficiency and
breakdown point (Donoho and Huber 1983) are two traditionally used
important criteria to compare different robust methods. The
efficiency is used to measure the relative efficiency of the robust
estimate compared to the OLS estimate when the error distribution is
exactly normal and there are no outliers. Breakdown point is to
measure the proportion of outliers an estimate can tolerate before
it goes to infinity. In this paper, finite sample breakdown point
(Donoho and Huber 1983) is used and defined as follows: Let
$\textbf{z}_{i} = \left(\bx_{i},y_{i}\right)$. Given any sample $\bz
= \left(\bz_{i},\ldots,\bz_{n}\right)$,  denote $T(\bz)$ the
estimate of the parameter $\bbeta$. Let $\bz'$ be the corrupted
sample where any $m$ of the original points of $\bz$ are replaced by
arbitrary bad data. Then the finite sample breakdown point
$\delta^{*}$ is defined as
\begin{equation}
\delta^{*}\left(\bz,T\right)=\min_{1\leq m\leq n}\left\{\frac{m}{n}:
\sup_{\bz'}\left\|T\left(\bz'\right)- T\left( \bz\right)\right\| =
\infty\right\},
\end{equation}
where $\left\|\cdot\right\|$ is Euclidean norm.

Many robust methods have been proposed to achieve high breakdown
point or high efficiency or both. M-estimates (Huber, 1981) are
solutions of the normal equation with appropriate weight functions.
They are resistant to unusual $\textit{y}$ observations, but
sensitive to high leverage points on $\textbf{x}$. Hence the
breakdown point of an M-estimate is $1/n$. R-estimates (Jaeckel
1972) which minimize the sum of scores of the ranked residuals have
relatively high efficiency but their breakdown points are as low as
those of OLS estimates. Least Median of Squares (LMS) estimates
(Siegel 1982) which minimize the median of squared residuals, Least
Trimmed Squares (LTS) estimates (Rousseeuw 1983) which minimize the
trimmed sum of squared residuals, and S-estimates (Rousseeuw and
Yohai 1984) which minimize the variance of the residuals all have
high breakdown point but with low efficiency. Generalized
S-estimates (GS-estimates) (Croux et al. 1994) maintain high
breakdown point as S-estimates and have slightly higher efficiency.
MM-estimates proposed by Yohai (1987) can simultaneously attain high
breakdown point and efficiencies. Mallows Generalized M-estimates
(Mallows 1975) and Schweppe Generalized M-estimates (Handschin et
al. 1975) downweight the high leverage points on $\textbf{x}$ but
cannot distinguish ``good" and ``bad" leverage points, thus
resulting in a loss of efficiencies. In addition, these two
estimators have low breakdown points when $\textit{p}$, the number
of explanatory variables, is large. Schweppe one-step (S1S)
Generalized M-estimates (Coakley and Hettmansperger 1993) overcome
the problems of Schweppe Generalized M-estimates and are calculated
in one step. They both have high breakdown points and high
efficiencies. Recently, Gervini and Yohai (2002) proposed a new
class of high breakdown point and high efficiency robust estimate
called robust and efficient weighted least squares estimator
(REWLSE). Lee et al. (2011) and She and Owen (2011) proposed a new
class of robust methods based on the regularization of case-specific
parameters for each response. They further proved that the
M-estimator with Huber's $\psi$ function is a special case of their
proposed estimator.


The rest of the paper is organized as follows. In Section 2, we review and describe some of the available robust methods. In Section 3, a simulation study and a real data application are used to compare different robust methods. Some discussions are given in Section 4.



\section{Robust Regression Methods}

\subsection{M-Estimates}
By replacing the least squares criterion (\ref{lse}) with a robust criterion, M-estimate (Huber, 1964) of $\bbeta$ is
\begin{equation}
\hat{\bbeta}=\arg\min_{\bbeta}\sum_{i=1}^{n}\rho\left(\frac{y_i-\bx_i^T \bbeta}{\hat{\sigma}}\right),
\label{mest}
\end{equation}
where $\rho(\cdot)$ is a robust loss function and $\hat{\sigma}$ is
an error scale estimate.  The derivative of $\rho$, denoted by
$\psi(\cdot)=\rho'(\cdot)$, is called the influence function. In
particular, if $\rho(t)$ = $ \frac{1}{2}t^{2}$, then the solution is
the OLS estimate. The OLS estimate is very sensitive to outliers.
Rousseeuw and Yohai (1984) indicated that OLS estimates have a
breakdown point (BP) of BP = $1/n$, which tends to zero when the
sample size $n$ is getting large. Therefore, one single unusual
observation can have large impact on the OLS estimate.

One of the commonly used robust loss functions is Huber's $\psi$ function (Huber 1981), where $\psi_c(t)=\rho'(t)=\max\{-c,\min(c,t)\}$. Huber (1981) recommends using $c=1.345$ in practice. This choice produces
a relative efficiency of approximately $95\%$ when the error density is normal. Another possibility for $\psi(\cdot)$ is Tukey's bisquare function $\psi_c(t)=t\{1-(t/c)^2\}_{+}^2$. The use of $c=4.685$ produces $95\%$ efficiency.
 If $\rho(t)$ = $\left|t \right|$, then \emph{least absolute deviation} (LAD, also called median regression) estimates are achieved by minimizing the sum of the absolute values of the residuals
\begin{equation}
\hat{\bbeta}=\arg\min_{\bbeta}\sum_{i=1}^{n} \left|y_i-\bx_i^T\bbeta\right|.
\end{equation}
The LAD is also called $L_{1}$ estimate due to the $L_{1}$ norm  used. Although LAD is more resistent than OLS to unusual $y$ values, it is sensitive to high leverage outliers, and thus has a breakdown point of BP = $1/n$ $\rightarrow 0$ (Rousseeuw and Yohai 1984). Moreover, LAD estimates have a low efficiency of 0.64 when the errors are normally distributed.  Similar to LAD estimates, the general monotone M-estimates, i.e., M-estimates with monotone $\psi$ functions, have a BP = $1/n$ $\rightarrow 0$ due to lack of immunity to high leverage outliers (Maronna, Martin, and Yohai 2006).


\subsection{LMS Estimates}
The LMS estimates (Siegel 1982) are found by minimizing the median of the squared residuals
\begin{equation}
\hat{\bbeta}=\arg\min_{\bbeta}\text{Med}\{\left( y_i-\bx_i^T\bbeta\right)^{2}\}.
\end{equation}
 One good property of the LMS estimate is that it possesses a high breakdown point of near 0.5. However, the LMS estimate has at best an efficiency of 0.37 when the assumption of normal errors is met (see Rousseeuw and Croux 1993). Moreover, LMS estimates do not have a well-defined influence function because of its convergence rate of $n^{-\frac{1}{3}}$ (Rousseeuw 1982). Despite these limitations, the LMS estimate can be used as the initial estimate for some other high breakdown point and  high efficiency robust methods.

\subsection{LTS Estimates}
The LTS estimate (Rousseeuw 1983) is defined as
\begin{equation}
\hat{\bbeta}=\arg\min_{\bbeta}\sum_{i=1}^{q}r_{(i)}\left( \bbeta\right)^{2},
\end{equation}
where $r_{(i)}(\bbeta)=y_{(i)}-\bx_{(i)}^T\bbeta$, $r_{\left(1\right)}\left(\bbeta\right)^{2} \leq\cdots\leq r_{\left(q\right)}\left(\bbeta\right)^{2}$ are ordered squared residuals, $q = \left[n \left(1-\alpha \right)+1 \right]$, and $\alpha$ is the proportion of trimming. Using q = $\left( \frac {n}{2} \right)$ +1 ensures that the estimator has a breakdown point of BP $= 0.5$, and the convergence rate of $n^{-\frac{1}{2}}$ (Rousseeuw 1983). Although highly resistent to outliers, LTS suffers badly in terms of very low efficiency, which is about 0.08, relative to OLS estimates (Stromberg, et al. 2000). The reason that LTS estimates call attentions to us is that it is traditionally used as the initial estimate for some other  high breakdown point and  high efficiency robust methods.

\subsection{S-Estimates}
S-estimates (Rousseeuw and Yohai 1984) are defined by
\begin{equation}
\hat{\bbeta}=\arg\min_{\bbeta}\hat{\sigma}\left(r_{1}\left(\bbeta\right),\cdots,r_{n}\left(\bbeta\right)\right),
\end{equation}
where $r_{i}\left(\bbeta\right)=y_i-\bx_i^T\bbeta$ and $\hat{\sigma}\left(r_{1}\left(\bbeta\right),\cdots,r_{n}\left(\bbeta\right)\right)$ is the scale M-estimate which is defined as the solution of
\begin{equation}
\frac{1}{n}\sum_{i=1}^{n}\rho \left(\frac{r_{i}\left( \bbeta \right)}{\hat{\sigma}}\right)=\delta,
\end{equation}
for any given $\bbeta$, where $\delta$ is taken to be $\e_{\Phi}\left[\rho\left(r\right)\right]$. For the biweight scale, S-estimates can attain a high breakdown point of BP = 0.5 and has an asymptotic efficiency of 0.29 under the assumption of normally distributed errors (Maronna, Martin, and Yahai 2006).

\subsection{Generalized S-Estimates (GS-Estimates)}
Croux et al. (1994) proposed generalized S-estimates in an attempt to improve the low efficiency of S-estimators. Generalized S-estimates are defined as
\begin{equation}
\hat{\bbeta} = \arg\min_{\bbeta}S_{n}(\bbeta),
\end{equation}
where $S_{n}(\bbeta)$ is defined as
\begin{equation}
S_{n}(\bbeta)=\sup\left\{S>0;
\binom{n}{2}^{-1}\sum_{i<j}\rho\left(\frac{r_{i}-r_{j}}{S}\right)
\geq k_{n,p}\right\},
\end{equation}
where $r_{i} = y_{i} - \bx_{i}^T\bbeta$, $\textit{p}$ is the number
of regression parameters, and $k_{n,p}$ is a constant which might
depend on $n$ and $p$. Particularly, if $\rho(x)=
I(\left|x\right|\geq 1)$ and $k_{n,p}  =
\left(\binom{n}{2}-\binom{h_{p}}{2}+1\right)/\binom{n}{2}$ with
$h_{p} = \frac{n+p+1}{2}$,  generalized S-estimator yields a special
case, the least quartile difference (LQD) estimator, which is
defined as
\begin{equation}
\hat{\bbeta} = \arg\min_{\bbeta}Q_{n}(r_{1},\ldots,r_{n}),
\end{equation}
where
\begin{equation}
Q_{n} =\left\{\left|r_{i}-r_{j}\right|;
i<j\right\}_{\binom{h_{p}}{2}}
\end{equation}
is the $\binom{h_{p}}{2}$th order statistic among the $\binom{n}{2}$
elements of the set $\left\{\left|r_{i}-r_{j}\right|; i<j\right\}$.
Generalized S-estimates have a breakdown point as high  as
S-estimates but with a higher efficiency.

\subsection{MM-Estimates}
First proposed by Yohai (1987), MM-estimates have become increasingly popular and are one of the most commonly employed robust regression techniques. The MM-estimates can be found by a three-stage procedure. In the first stage, compute an initial consistent estimate $\hat{\bbeta}_{0}$ with high breakdown point but possibly low normal efficiency. In the second stage, compute a robust M-estimate of scale $\hat{\sigma}$ of the residuals based on the initial estimate. In the third stage, find an M-estimate $\hat{\bbeta}$ starting at $\hat{\bbeta}_{0}$.

In practice, LMS or S-estimate with Huber or bisquare functions is typically used as the initial estimate $\hat{\bbeta_{0}}$.
Let $\rho_{0} (r) =\rho_{1}\left( r/k_{0}\right)$, $\rho (r) =\rho_{1}\left( r/k_{1}\right)$, and assume that each of the $\rho$-functions is bounded. The scale estimate $\hat{\sigma}$ satisfies
\begin{equation}
\frac{1}{n}\sum_{i=1}^{n}\rho_{0} \left(\frac{r_{i}\left( \hat{\bbeta} \right)}{\hat{\sigma}}\right)=0.5.
\end{equation}
If the $\rho$-function is biweight, then $k_{0} = 1.56$ ensures that the estimator has the asymptotic BP = 0.5. Note that an M-estimate minimizes
\begin{equation}
L(\beta) = \sum_{i=1}^{n}\rho \left(\frac{r_{i}\left( \hat{\bbeta} \right)}{\hat{\sigma}}\right).
\end{equation}
Let $\rho$ satisfy $\rho\leq \rho_{0}$. Yohai (1987) showed that if $\hat{\bbeta}$ satisfies $L(\hat{\bbeta})\leq (\hat{\bbeta}_{0})$, then $\hat{\bbeta}$'s BP is not less than that of $\hat{\bbeta}_{0}$. Furthermore, the breakdown point of the MM-estimate depends only on $k_{0}$ and the asymptotic variance of the MM-estimate depends only on $k_{1}$. We can choose $k_{1}$ in order to attain the desired normal efficiency without affecting its breakdown point. In order to let $\rho\leq \rho_{0}$, we must have $k_{1} \geq k_{0}$; the larger the $k_{1}$ is, the higher efficiency the MM-estimate can attain at the normal distribution.

Maronna, Martin, and Yahai (2006) provides the values of $k_{1}$ with the corresponding efficiencies of the biweight $\rho$-function. Please see the following table for more detail.
\begin{center}
\begin{tabular}{c c c c c} \hline
Efficiency &   0.80   & 0.85 & 0.90 & 0.95 \\
\hline
$k_{1}$ & 3.14 & 3.44 &3.88 & 4.68\\
\hline
\end{tabular}
\end{center}
However, Yohai (1987) indicates that MM-estimates with larger values of $k_{1}$ are more sensitive to outliers than the estimates corresponding to smaller values of $k_{1}$. In practice, an MM-estimate with bisquare function and efficiency 0.85 ($k_{1}$ = 3.44) starting from a bisquare S-estimate is recommended. 

\subsection{Generalized M-Estimates (GM-Estimates)}

\subsubsection{Mallows GM-estimate}
In order to make M-estimate resistent to high leverage outliers, Mallows (1975) proposed Mallows GM-estimate that is defined by
\begin{equation}
\sum_{i=1}^{n}w_{i}\psi\left\{\frac{r_{i}\left(\hat{\bbeta}\right)}{\hat{\sigma}}\right\}\bx_{i}=0,
\end{equation}
where $\psi(e)=\rho'(e)$ and $w_{i} = \sqrt{1-h_{i}}$ with $h_{i}$ being the leverage of the $\textit{i}th$ observation. 
The weight $w_{i}$ ensures that the observation with high leverage
receives less weight than observation with small leverage. However,
even ``good" leverage points that fall in line with the pattern in
the bulk of the data are down-weighted, resulting in a loss of
effiency.

\subsubsection{Schweppe GM-estimate}
Schweppe GM-estimate (Handschin et al. 1975) is defined by the solution of
\begin{equation}
\sum_{i=1}^{n}w_{i}\psi\left\{\frac{r_{i}\left(\hat{\bbeta}\right)}{w_{i}\hat{\sigma}}\right\}\bx_{i}=0,
\end{equation}
which adjusts the leverage weights according to the size of the residual $r_{i}$. Carroll and Welsh (1988) proved that the Schweppe estimator is not consistent when the errors are asymmetric. Furthermore, the breakdown points for both Mallows and Schweppe GM-estimates are no more than $1/(p+1)$, where $\textit{p}$ is the number of unknown parameters.

\subsubsection{S1S GM-estimate}
Coakley and Hettmansperger (1993) proposed Schweppe one-step (S1S) estimate , which extends from the original Schweppe estimator. 
S1S estimator is defined as
\begin{equation}
\hat{\bbeta}= \hat{\bbeta}_{0}+ \left[\sum_{i=1}^{n}\psi'\left(\frac{r_{i}\left(\hat{\bbeta}_{0}\right)}{\hat{\sigma}w_{i}}\right)\bx_{i}\bx_{i}'\right]^{-1}\times \sum_{i=1}^{n}\hat{\sigma}w_{i}\psi\left(\frac{r_{i}\left(\hat{\bbeta}_{0}\right)}{\hat{\sigma}w_{i}}\right)\bx_{i},
\end{equation}
where the weight $w_{i}$ is defined in the same way as Schweppe's GM-estimate.

The method for S1S estimate is different from the Mallows and Schweppe GM-estimates in that once the initial estimates of the residuals and the scale of the residuals are given, final M-estimates are calculated in one step rather than iteratively. Coakley and Hettmansperger (1993) recommended to use Rousseeuw's LTS for the initial estimates of the residuals and LMS for the initial estimates of the scale and proved that the S1S estimate gives a breakdown point of BP = 0.5 and results in 0.95 efficiency compared to the OLS estimate under the Gauss-Markov assumption.

\subsection{R-Estimates}
The R-estimate (Jaeckel 1972) minimizes the sum of some scores of the ranked residuals
\begin{equation}
\sum_{i=1}^{n}a_{n}\left(R_{i} \right)r_{i}=min,
\end{equation}
where $R_{i}$ represents the rank of the \textit{i}th residual $r_{i}$, and  $a_{n}\left(\cdot\right)$ is a monotone score function that satisfies
\begin{equation}
\sum_{i=1}^{n}a_{n}\left(i\right) = 0.
\end{equation}
R-estimates are scale equivalent which is an advantage compared to M-estimates. However, the optimal choice of the score function is unclear. In addition, most of R-estimates have a breakdown point of BP = $1/n \rightarrow 0$. The bounded influence R-estimator proposed by Naranjo and Hettmansperger (1994) has a fairly high efficiency when the errors have normal distribution. However, it is proved that their breakdown point is no more than 0.2.

\subsection{REWLSE}
Gervini and Yohai (2002) proposed a new class of robust regression method called robust and efficient weighted least squares estimator (REWLSE). REWLSE is much more attractive than many other robust estimators due to its simultaneously attaining maximum breakdown point and full efficiency under normal errors. This new estimator is a type of weighted least squares estimator with the weights adaptively calculated from an initial robust estimator.

Consider a pair of initial robust estimates of regression parameters and scale, $\hat{\bbeta}_0$ and $\hat{\sigma}$ respectively, the standardized residuals are defined as
\[
r_{i}=\frac{y_{i}-\bx_{i}^T\hat{\bbeta}_{0}}{\hat{\sigma}}.
\]
A large value of $\left|r_{i}\right|$ would suggest that $(\bx_{i},y_{i})$ is an outlier.
Define a measure of proportion of outliers in the sample
\begin{equation}
d_{n} = \max_{i>i_{0}}\left\{F^{+}(\left|r\right|_{(i)})-\frac{(i-1)}{n}\right\}^{+},
\end{equation}
where $\left\{\cdot\right\}^{+}$ denotes positive part, $F^{+}$
denotes the distribution of $\left|X\right|$ when $X \thicksim F$,
$\left|r\right|_{(1)} \leq \ldots \leq \left|r\right|_{(n)}$ are the
order statistics of the standardized absolute residuals, and $i_{0}
= \max\left\{i: \left|r\right|_{(i)} < \eta\right\}$, where $\eta$
is some large quantile of $F^{+}$.  Typically $\eta = 2.5$ and  the
cdf of a normal distribution is  chosen for $F$. Thus those
$\left\lfloor nd_{n}\right\rfloor$ observations with largest
standardized absolute residuals are eliminated (here $\left\lfloor
a\right\rfloor$ is the largest integer less than or equal to a).

The adaptive cut-off value is
$t_{n} = \left|r\right|_{(i_{n})}$ with $i_{n} = n - \left\lfloor nd_{n}\right\rfloor$.
With this adaptive cut-off value, the adaptive weights proposed by Gervini and Yohai (2002) are
\begin{equation}
w_{i}=
\begin{cases}
1   &\text{if $\left|r_{i}\right| < t_{n}$}\\
0   &\text{if $\left|r_{i}\right| \geq t_{n}.$}
\end{cases}
\end{equation}
Then, the REWLSE is
\begin{equation}
\hat{\bbeta} =
(\bX^T\bW\bX)^{-1}\bX^T\bW\by,   
\end{equation}
where $\bW =\diag(w_{1},\cdots, w_{n}),\bX=(\bx_1,\ldots,\bx_n)^T,$
and $\textbf{y}=(y_{1},\cdots, y_{n})'$.

If the initial regression and scale estimates with BP = 0.5 are
chosen, the breakdown point of the REWLSE is also 0.5. Furthermore,
when the errors are normally distributed, the REWLSE is
asymptotically equivalent to the OLS estimates and hence
asymptotically efficient.

\subsection{Robust regression based on regularization of case-specific parameters}
She and Owen (2011) and Lee et al$.$ (2011) proposed a new class of robust regression methods using the case-specific indicators in a mean shift model with regularization method. A mean shift model for the linear regression is
\[
\by=\bX\bbeta + \bgamma +\bvarepsilon, \; \bvarepsilon \thicksim N(0,\sigma^{2}I)
\]
where $\textbf{y}=(y_{1},\cdots, y_{n})^T$, $\bX=(\bx_1,\ldots,\bx_n)^T$, and the mean shift parameter $\gamma_{i}$ is nonzero when the \textit{i}th observation is an outlier and zero, otherwise.

Due to the sparsity of $\gamma_i$s, She and Owen (2011) and Lee et al$.$ (2011) proposed to estimate $\bbeta$ and $\bgamma$ by minimizing the penalized least squares using $L_1$ penalty:
\begin{equation}
L(\bbeta,\bgamma) = \frac{1}{2}\left\{\by-(\bX\bbeta+\bgamma)\right\}^{T}\left\{\by-(\bX\bbeta+\bgamma)\right\}+\lambda\sum_{i=1}^{n}\left|\gamma_{i}\right|,
\label{penlse}
\end{equation}
where $\lambda$ are fixed regularization parameters for $\bgamma$. Given the estimate $\hat{\bgamma}$, $\hat{\bbeta}$ is the OLS estimate with $\textbf{y}$ replaced by $\textbf{y}-\bgamma$. For a fixed $\hat{\bbeta}$, the minimizer of (\ref{penlse}) is $\hat{\gamma}_i= sgn(r_i)(\left|\gamma_i\right|-\lambda)_{+}$, that is,
\[
\hat{\gamma}_i=
\begin{cases}
0                               &\text{if $\left|r_{i}\right|\leq \lambda$;}\\
y_{i}-\bx_{i}^{T}\hat{\bbeta}      &\text{if $\left|r_{i}\right|> \lambda$.}
\end{cases}
\]
Therefore, the solution of (\ref{penlse}) can be found by iteratively updating the above two steps. She and Owen (2011) and Lee et al$.$ (2011) proved that the above estimate is in fact equivalent to the M-estimate if Huber's $\psi$ function is used. However, their proposed robust estimates are based on different perspective and can be extended to many other likelihood based models.

Note, however, the monotone M-estimate is not resistent to the high leverage outliers. In order to overcome this problem, She and Owen (2011) further proposed to replace the $L_1$ penalty in (\ref{penlse}) by a general penalty. The objective function is then defined by
\begin{equation}
L_{p}(\bbeta,\bgamma) = \frac{1}{2}\left\{\by-(\bX\bbeta+\bgamma)\right\}^{T}\left\{\by-(\bX\bbeta+\bgamma)\right\}+\sum_{i=1}^{n}p_{\lambda}(\left|\gamma_{i}\right|),
\label{lsepen}
\end{equation}
where $p_{\lambda}(\left|\cdot\right|)$ is any penalty function which depends on the regularization parameter $\lambda$. We can find $\hat{\bgamma}$ by defining thresholding function $\Theta(\bgamma;\lambda)$ (She and Owen 2009). She and Owen (2009, 2011) proved that for a specific thresholding function, we can always find the corresponding penalty function. For example, the soft, hard, and smoothly clipped absolute deviation (SCAD; Fan and Li, 2001) thresholding solutions of $\bgamma$ correspond to  $L_{1}$, Hard, and SCAD penalty functions, respectively. Minimizing the equation (\ref{lsepen}) yields a sparse $\hat{\bgamma}$ for outlier detection and a robust estimate of $\bbeta$. She and Owen (2011) showed that
the proposed estimates of (\ref{lsepen}) with hard or SCAD penalties are equivalent to the M-estimates with certain redescending $\psi$ functions and thus will be resistent to high leverage outliers if a high breakdown point robust estimates are used as the initial values.


\section{Examples}
In this section, we compare different robust methods and report the
mean squared errors (MSE) of the parameter estimates for each
estimation method. We compare the OLS estimate with seven other
commonly used robust regression estimates: the M estimate using
Huber's $\psi$ function ($M_H$), the M estimate using Tukey's
bisquare function ($M_T$), the S estimate, the LTS estimate, the LMS
estimate, the MM estimate (using bisquare weights and $k_{1} =
4.68$), and the REWLSE. Note that we didn't include the
case-specific regularization methods proposed by She and Owen (2011)
and Lee et al$.$ (2011) since they are essentially equivalent to
M-estimators (She and Owen (2011) did show that their new methods
have better performance in detecting outliers in their simulation
study).



\bigskip

{\bf  Example 1.} We generate $n$ samples
$\{(x_1,y_1),\ldots,(x_n,y_n)\}$ from the model \[Y = X +
\varepsilon,\] where $X \sim N (0,1)$.
In order to compare the performance of different methods, we consider the following six cases for the error density of $\varepsilon$:
\begin{description}
\item[Case I:] $\varepsilon\sim N (0,1)$- standard normal distribution.
\item[Case II:] $\varepsilon\sim t_{3}$ - t-distribution with degrees of freedom 3.
\item[Case III:] $\varepsilon\sim t_{1}$ - t-distribution with degrees of freedom 1 (Cauchy distribution).
\item[Case IV:] $\varepsilon\sim 0.95N (0,1) + 0.05N (0,10^{2})$ - contaminated normal mixture.
\item[Case V:] $\varepsilon\sim$ N (0,1) with $10\%$ identical outliers in $y$ direction (where we let the first $10\%$ of $y's$ equal to 30).
\item[Case VI:] $\varepsilon\sim$ N (0,1) with $10\%$ identical high leverage outliers (where we let the first $10\%$ of $x's$ equal to 10 and their corresponding $y's$ equal to 50).
\end{description}

Tables 1 and 2 report the mean squared errors (MSE) of the parameter
estimates for each estimation method with sample size $n = 20$ and
100, respectively. The number of replicates is 200.  From the
tables, we can see that MM and REWLSE have the overall best
performance throughout most cases and they are consistent for
different sample sizes. For Case I, LSE has the smallest MSE which
is reasonable since under normal errors LSE is the best estimate;
$M_H$, $M_T$, MM, and REWLSE have similar MSE to LSE, due to their
high efficiency property; LMS, LTS, and S have relative larger MSE
due to their low efficiency. For Case II, $M_H$, $M_T$, MM, and
REWLSE work better than other estimates. For Case III, LSE has much
larger MSE than other robust estimators; $M_H$, $M_T$, MM, and
REWLSE have similar MSE to S. For Case IV, M, MM, and REWLSE have
smaller MSE than others. From Case V, we can see that when the data
contain outliers in the y-direction, LSE is much worse than any
other robust estimates; MM, REWLSE, and $M_T$ are better than other
robust estimators. Finally for Case VI, since there are high
leverage outliers, similar to LSE, both $M_T$ and $M_H$ perform
poorly; MM and REWLSE work better than other robust estimates.

In order to better compare the performance of different methods,
Figure 1 shows the plot of their MSE versus each case for the slope
(left side) and intercept (right side) parameters for model 1 when
sample size $n=100$. Since the lines for LTS and LMS are above the
other lines, S, MM, and REWLSE of the intercept and slopes
outperform LTS and LMS estimates throughout all six cases. In
addition, the S estimate has similar performance to MM and REWLSE
when the error density of $\varepsilon$ is Cauchy distribution.
However, MM and REWLSE perform better than S-estimates in other five
cases. Furthermore, the lines for MM and REWLSE almost overlap for
all six cases. It shows that MM and REWLSE are the overall best
approaches in robust regression.

\bigskip

{\bf Example 2.} \[Y=X_{1}+X_{2}+X_{3}+\varepsilon,\] where $X_i\sim
N(0,1), i=1,2,3$ and $X_i$'s are independent. We consider the
following six cases for the error density of $\varepsilon$:
\begin{description}
\item[Case I:] $\varepsilon\sim N (0,1)$- standard normal distribution.
\item[Case II:] $\varepsilon\sim t_{3}$ - t-distribution with degrees of freedom 3.
\item[Case III:] $\varepsilon\sim t_{1}$ - t-distribution with degrees of freedom 1 (Cauchy distribution).
\item[Case IV:] $\varepsilon\sim 0.95N (0,1) + 0.05N (0,10^{2})$ - contaminated normal mixture.
\item[Case V:] $\varepsilon\sim N (0,1)$ with $10\%$ identical outliers in $y$ direction (where we let the first $10\%$ of $y's$ equal to 30).
\item[Case VI:] $\varepsilon\sim N (0,1)$ with $10\%$ identical high leverage outliers (where we let the first $10\%$ of $x's$ equal to 10 and their corresponding $y's$ equal to 50).
\end{description}

Tables 3 and 4 show the mean squared errors (MSE) of the parameter
estimates of each estimation method for sample size $n = 20$ and $n
= 100$, respectively. Figure 2 shows the plot of their MSE versus
each case for three slopes and the intercept parameters with sample
size $n =100$. The results in Example 2 tell similar stories to
Example 1. In summary, MM and REWLSE have the overall best
performance; LSE only works well when there are no outliers since it
is very sensitive to outliers; M-estimates ($M_H$ and $M_T$) work
well if the outliers are in $y$ direction but are also sensitive to
the high leverage outliers.

\bigskip

\textbf{Example 3:}  Next, we use the famous data set found in
Freedman et al$.$ (1991) to compare LSE with MM and REWLSE. The data
set are shown in Table 5 which contains per capita consumption of
cigarettes in various countries in 1930 and the death rates (number
of deaths per million people) from lung cancer for 1950. Here, we
are interested in how the death rates per million people from lung
cancer (dependent variable $y$)  dependent on the consumption of
cigarettes per capita (the independent variable $x$). Figure
\ref{figure3} is a scatter plot of the data. From the plot, we can
see that USA $(x=1300,y=200)$ is an outlier with high leverage. We
compare different regression parameters estimates by LSE, MM, and
REWLSE. Figure \ref{figure3} shows the fitted lines by these three
estimates. The LSE line does not fit the bulk of the data, being a
compromise between USA observation and the rest of the data, while
the fitted lines for the other two estimates almost overlap and give
a better representation of the majority of the data.

Table 6 also gives the estimated regression parameters of these
three methods for both the complete data and the data without the
outlier USA. For LSE, the intercept estimate changes from 67.56
(complete data set) to 9.14 (without outlier) and the slope estimate
changes from 0.23 (complete data set) to 0.37 (without outlier).
Thus, it is clear that the outlier USA strongly influences LSE. For
MM-estimate, after deleting the outlier, the intercept estimate
changes slightly but slope estimate remains almost the same. For
REWLSE, both intercept and slope estimates remain unchanged after
deleting the outlier. In addition, note that REWLSE for the whole
data gives almost the same result as LSE without the outlier.

\section{Discussion}
In this article, we describe and compare different available robust
methods. Table 7 summarizes the robustness attributes and asymptotic
efficiency of most of the estimators we have discussed. Based on
Table 7, it can be seen that MM-estimates and REWLSE have both high
breakdown point and high efficiency. Our simulation study also
demonstrated that MM-estimates and REWLSE have overall best
performance among all compared robust methods. In terms of breakdown
point and efficiency, GM-estimates (Mallows, Schweppe), Bounded
R-estimates, M-estimates, and LAD estimates are less attractive due
to their low breakdown points. Although LMS, LTS, S-estimates, and
GS-estimates are strongly resistent to outliers, their efficiencies
are low. However, these high breakdown point robust estimates such
as S-estimates and LTS are traditionally used as the initial
estimates for some other high breakdown point and high efficiency
robust estimates.



\begin{table}[htb]
\label{tab1} \centering\caption{MSE of Point Estimates for Example 1 with $n = 20$} \vskip 0.05in
\def\arraystretch{1}
\small \hspace*{-22.75pt}
\begin{tabular}{c|c c c c c c c c} \hline
 TRUE &  OLS  & $M_{H}$ & $M_{T}$ & LMS & LTS & S &  MM & REWLSE\\
\hline
& \multicolumn{7}{c}{Case I: $\varepsilon\sim N (0,1)$}\\
$\beta_{0}: 0$ & 0.0497 & 0.0532 & 0.0551 & 0.2485 & 0.2342 & 0.1372 & 0.0564 & 0.0645\\
$\beta_{1}: 1$ & 0.0556 & 0.0597 & 0.0606 & 0.2553 & 0.2328 & 0.1679 & 0.0643 & 0.0733\\
\hline
& \multicolumn{7}{c}{ Case II: $\varepsilon\sim t_{3}$}\\
$\beta_{0}: 0$ & 0.1692 & 0.0884 & 0.0890 & 0.3289 & 0.3076 & 0.1637 & 0.0856 & 0.0982\\
$\beta_{1}: 1$ & 0.1766 & 0.1041 & 0.1027 & 0.4317 & 0.3905 & 0.2041 & 0.1027 & 0.1189\\
\hline
&\multicolumn{7}{c}{Case III: $\varepsilon\sim t_{1}$}\\
$\beta_{0}: 0$ & 1003.8360 & 0.2545 & 0.2146 & 0.3215 & 0.2872 & 0.1447 & 0.1824 & 0.1990\\
$\beta_{1}: 1$ & 1374.0645 & 0.4103 & 0.3209 & 0.3659 & 0.3496 & 0.1843 & 0.2996 & 0.3164\\
\hline
& \multicolumn{7}{c}{ Case IV: $\varepsilon\sim 0.95N (0,1) + 0.05N (0,10^{2})$}\\
$\beta_{0}: 0$ & 0.3338 & 0.0610 & 0.0528 & 0.2105 & 0.2135 & 0.1228 & 0.0523 & 0.0538\\
$\beta_{1}: 1$ & 0.4304 & 0.0808 & 0.0644 & 0.3149 & 0.2908 & 0.1519 & 0.0636 & 0.0691\\
\hline
& \multicolumn{7}{c}{ Case V: $\varepsilon\sim N (0,1)$ with outliers in $y$ direction}\\
$\beta_{0}: 0$ & 9.3051 & 0.1082 & 0.0697 & 0.2752 & 0.2460 & 0.1430 & 0.0671 & 0.0667\\
$\beta_{1}: 1$ & 5.5747 & 0.1083 & 0.0762 & 0.2608 & 0.2029 & 0.1552 & 0.0746 & 0.0801\\
\hline
& \multicolumn{7}{c}{ Case VI: $\varepsilon\sim N (0,1)$ with high leverage outliers}\\
$\beta_{0}: 0$ & 0.8045 & 0.8711 & 0.8857 & 0.2161 & 0.1984 & 0.1256 & 0.0581 & 0.0598\\
$\beta_{1}: 1$ & 13.4258 & 13.7499 & 13.8487 & 0.3377 & 0.3019 & 0.1695 & 0.0749 & 0.0749\\
\hline

\end{tabular}
\end{table}

\begin{table}[htb]
\label{tab2} \centering\caption{MSE of Point Estimates for Example 1
with $n = 100$} \vskip 0.05in
\def\arraystretch{1}
\small \hspace*{-22.75pt}
\begin{tabular}{c|c c c c c c c c} \hline
 TRUE &  OLS  &  $M_{H}$ & $M_{T}$ & LMS & LTS & S &  MM & REWLSE\\
\hline
& \multicolumn{7}{c}{Case I: $\varepsilon\sim N (0,1)$}\\
$\beta_{0}: 0$ & 0.0113 & 0.0126 & 0.0125 & 0.0755 & 0.0767 & 0.0347 & 0.0125 & 0.0131\\
$\beta_{1}: 1$ & 0.0096 & 0.0102 & 0.0103 & 0.0693 & 0.0705 & 0.0312 & 0.0103 & 0.0112\\
\hline
& \multicolumn{7}{c}{ Case II: $\varepsilon\sim t_{3}$}\\
$\beta_{0}: 0$ & 0.0283 & 0.0154 & 0.0153 & 0.0596 & 0.0659 & 0.0231 & 0.0153 & 0.0170\\
$\beta_{1}: 1$ & 0.0255 & 0.0157 & 0.0164 & 0.0652 & 0.0752 & 0.0356 & 0.0163 & 0.0185\\
\hline
&\multicolumn{7}{c}{Case III: $\varepsilon\sim t_{1}$}\\
$\beta_{0}: 0$ & 40.8454 & 0.0416 & 0.0310 & 0.0550 & 0.0392 & 0.0201 & 0.0323 & 0.0354\\
$\beta_{1}: 1$ & 39.5950 & 0.0469 & 0.0387 & 0.0607 & 0.0476 & 0.0274 & 0.0402 & 0.0447\\
\hline
& \multicolumn{7}{c}{ Case IV: $\varepsilon\sim 0.95N (0,1) + 0.05N (0,10^{2})$}\\
$\beta_{0}: 0$ & 0.0650 & 0.0119 & 0.0107 & 0.0732 & 0.0737 & 0.0296 & 0.0107 & 0.0110\\
$\beta_{1}: 1$ & 0.0596 & 0.0126 & 0.0123 & 0.0696 & 0.0775 & 0.0353 & 0.0122 & 0.0134\\
\hline
& \multicolumn{7}{c}{ Case V: $\varepsilon\sim N (0,1)$ with outliers in $y$ direction}\\
$\beta_{0}: 0$ & 8.9470 & 0.0465 & 0.0107 & 0.0674 & 0.0658 & 0.0283 & 0.0106 & 0.0108\\
$\beta_{1}: 1$ & 0.7643 & 0.0146 & 0.0120 & 0.0611 & 0.0704 & 0.0338 & 0.0119 & 0.0120\\
\hline
& \multicolumn{7}{c}{ Case VI: $\varepsilon\sim N (0,1)$ with high leverage outliers}\\
$\beta_{0}: 0$ & 0.2840 & 0.2999 & 0.2983 & 0.0575 & 0.0595 & 0.0234 & 0.0107 & 0.0106\\
$\beta_{1}: 1$ & 13.2298 & 13.5907 & 13.7210 & 0.0624 & 0.0790 & 0.0310 & 0.0127 & 0.0131\\
\hline
\end{tabular}
\end{table}

\begin{table}[htb]
\label{tab3} \centering\caption{MSE of Point Estimates for Example 2
with $n = 20$} \vskip 0.05in
\def\arraystretch{1}
\small \hspace*{-22.75pt}
\begin{tabular}{c|c c c c c c c c} \hline
 TRUE &  OLS  &  $M_{H}$ & $M_{T}$ & LMS & LTS & S &  MM & REWLSE\\
\hline
& \multicolumn{7}{c}{Case I: $\varepsilon\sim N (0,1)$}\\
$\beta_{0}: 0$ & 0.0610 & 0.0659 & 0.0744 & 0.3472 & 0.2424 & 0.1738 & 0.0679 & 0.0800\\
$\beta_{1}: 1$ & 0.0588 & 0.0664 & 0.0752 & 0.4066 & 0.3247 & 0.2299 & 0.0709 & 0.1051\\
$\beta_{2}: 1$ & 0.0620 & 0.0653 & 0.0725 & 0.3557 & 0.2724 & 0.2018 & 0.0716 & 0.0880\\
$\beta_{3}: 1$ & 0.0698 & 0.0719 & 0.0758 & 0.3444 & 0.2657 & 0.1904 & 0.0751 & 0.0999\\
\hline
& \multicolumn{7}{c}{ Case II: $\varepsilon\sim t_{3}$}\\
$\beta_{0}: 0$ & 0.1745 & 0.1125 & 0.1168 & 0.3799 & 0.3040 & 0.2326 & 0.1177 & 0.1210\\
$\beta_{1}: 1$ & 0.1998 & 0.1332 & 0.1364 & 0.4402 & 0.3404 & 0.2539 & 0.1311 & 0.1485\\
$\beta_{2}: 1$ & 0.1704 & 0.1203 & 0.1272 & 0.4868 & 0.3831 & 0.2118 & 0.1242 & 0.1461\\
$\beta_{3}: 1$ & 0.2018 & 0.1520 & 0.1732 & 0.5687 & 0.4964 & 0.3145 & 0.1649 & 0.2049\\
&\multicolumn{7}{c}{Case III: $\varepsilon\sim t_{1}$}\\
$\beta_{0}: 0$ & 248.0170 & 0.3492 & 0.2579 & 0.7935 & 0.4657 & 0.3615 & 0.2630 & 0.2957\\
$\beta_{1}: 1$ & 209.8339 & 0.4503 & 0.3713 & 1.2482 & 0.9701 & 0.4355 & 0.3784 & 0.4443\\
$\beta_{2}: 1$ & 93.1344 & 0.4089 & 0.2936 & 1.0517 & 0.6203 & 0.5086 & 0.2965 & 0.3365\\
$\beta_{3}: 1$ & 374.7307 & 0.4387 & 0.3206 & 1.0829 & 0.7704 & 0.4717 & 0.3123 & 0.4023\\
\hline
& \multicolumn{7}{c}{ Case IV: $\varepsilon\sim 0.95N (0,1) + 0.05N (0,10^{2})$}\\
$\beta_{0}: 0$ & 0.3245 & 0.0853 & 0.0837 & 0.2820 & 0.2433 & 0.1873 & 0.0785 & 0.0924\\
$\beta_{1}: 1$ & 0.3391 & 0.1026 & 0.1001 & 0.4609 & 0.2875 & 0.2328 & 0.0996 & 0.1047\\
$\beta_{2}: 1$ & 0.3039 & 0.0898 & 0.0938 & 0.4077 & 0.3053 & 0.1887 & 0.0900 & 0.1170\\
$\beta_{3}: 1$ & 0.2618 & 0.0846 & 0.0941 & 0.4560 & 0.3023 & 0.2054 & 0.0900 & 0.1007\\
\hline
\hline
& \multicolumn{7}{c}{ Case V: $\varepsilon\sim N (0,1)$ with outliers in $y$ direction}\\
$\beta_{0}: 0$ & 9.9455 & 0.1442 & 0.0706 & 0.3127 & 0.2334 & 0.1759 & 0.0680 & 0.0713\\
$\beta_{1}: 1$ & 5.1353 & 0.1015 & 0.0636 & 0.3638 & 0.2769 & 0.1508 & 0.0617 & 0.0654\\
$\beta_{2}: 1$ & 5.1578 & 0.1245 & 0.0730 & 0.4647 & 0.2796 & 0.1759 & 0.0690 & 0.0722\\
$\beta_{3}: 1$ & 6.0662 & 0.1273 & 0.0612 & 0.3922 & 0.2733 & 0.1797 & 0.0597 & 0.0654\\
\hline
& \multicolumn{7}{c}{ Case VI: $\varepsilon\sim N (0,1)$ with high leverage outliers}\\
$\beta_{0}: 0$ & 1.0096 & 1.0733 & 1.1334 & 0.3339 & 0.2491 & 0.1716 & 0.0821 & 0.0840\\
$\beta_{1}: 1$ & 13.6630 & 14.0715 & 14.1688 & 0.4698 & 0.3126 & 0.2500 & 0.1467 & 0.1031\\
$\beta_{2}: 1$ & 0.9201 & 0.9684 & 1.0108 & 0.4088 & 0.2681 & 0.2064 & 0.0899 & 0.1088\\
$\beta_{3}: 1$ & 0.8538 & 0.9316 & 0.9937 & 0.4411 & 0.3373 & 0.2077 & 0.0709 & 0.0957\\
\hline
\end{tabular}
\end{table}

\begin{table}[htb]
\label{tab4} \centering\caption{MSE of Point Estimates for Example 2
with $n = 100$} \vskip 0.05in
\def\arraystretch{1}
\small \hspace*{-22.75pt}
\begin{tabular}{c|c c c c c c c c} \hline
 TRUE &  OLS  &  $M_{H}$ & $M_{T}$  & LMS & LTS & S &  MM & REWLSE\\
\hline
& \multicolumn{7}{c}{Case I: $\varepsilon\sim N (0,1)$}\\
$\beta_{0}: 0$ & 0.0097 & 0.0108 & 0.0109 & 0.0743 & 0.0690 & 0.0359 & 0.0108 & 0.0119\\
$\beta_{1}: 1$ & 0.0111 & 0.0120 & 0.0121 & 0.0736 & 0.0778 & 0.0399 & 0.0119 & 0.0130\\
$\beta_{2}: 1$ & 0.0100 & 0.0106 & 0.0107 & 0.0713 & 0.0715 & 0.0404 & 0.0107 & 0.0114\\
$\beta_{3}: 1$ & 0.0110 & 0.0116 & 0.0118 & 0.0662 & 0.0712 & 0.0388 & 0.0118 & 0.0121\\
\hline
& \multicolumn{7}{c}{ Case II: $\varepsilon\sim t_{3}$}\\
$\beta_{0}: 0$ & 0.0294 & 0.0145 & 0.0159 & 0.0713 & 0.0655 & 0.0330 & 0.0158 & 0.0179\\
$\beta_{1}: 1$ & 0.0464 & 0.0198 & 0.0180 & 0.0651 & 0.0674 & 0.0368 & 0.0181 & 0.0195\\
$\beta_{2}: 1$ & 0.0375 & 0.0183 & 0.0181 & 0.0727 & 0.0733 & 0.0352 & 0.0181 & 0.0195\\
$\beta_{3}: 1$ & 0.0365 & 0.0176 & 0.0167 & 0.0646 & 0.0736 & 0.0344 & 0.0167 & 0.0175\\
\hline
&\multicolumn{7}{c}{Case III: $\varepsilon\sim t_{1}$}\\
$\beta_{0}: 0$ & 36.7303 & 0.0388 & 0.0287 & 0.0681 & 0.0590 & 0.0317 & 0.0289 & 0.0326\\
$\beta_{1}: 1$ & 31.6433 & 0.0499 & 0.0351 & 0.0624 & 0.0618 & 0.0262 & 0.0367 & 0.0372\\
$\beta_{2}: 1$ & 41.4547 & 0.0422 & 0.0337 & 0.0788 & 0.0613 & 0.0321 & 0.0344 & 0.0369\\
$\beta_{3}: 1$ & 29.7017 & 0.0476 & 0.0317 & 0.0714 & 0.0506 & 0.0320 & 0.0332 & 0.0362\\
\hline
& \multicolumn{7}{c}{ Case IV: $\varepsilon\sim 0.95N (0,1) + 0.05N (0,10^{2})$}\\
$\beta_{0}: 0$ & 0.0591 & 0.0109 & 0.0100 & 0.0656 & 0.0625 & 0.0281 & 0.0100 & 0.0109\\
$\beta_{1}: 1$ & 0.0492 & 0.0122 & 0.0112 & 0.0558 & 0.0643 & 0.0349 & 0.0110 & 0.0115\\
$\beta_{2}: 1$ & 0.0640 & 0.0123 & 0.0110 & 0.0635 & 0.0683 & 0.0337 & 0.0109 & 0.0118\\
$\beta_{3}: 1$ & 0.0696 & 0.0135 & 0.0122 & 0.0573 & 0.0608 & 0.0333 & 0.0122 & 0.0128\\
\hline
& \multicolumn{7}{c}{ Case V: $\varepsilon\sim N (0,1)$ with outliers in $y$ direction}\\
$\beta_{0}: 0$ & 9.1058 & 0.0560 & 0.0118 & 0.0631 & 0.0579 & 0.0322 & 0.0118 & 0.0120\\
$\beta_{1}: 1$ & 0.8544 & 0.0186 & 0.0137 & 0.0738 & 0.0814 & 0.0377 & 0.0136 & 0.0143\\
$\beta_{2}: 1$ & 0.9538 & 0.0189 & 0.0141 & 0.0672 & 0.0717 & 0.0379 & 0.0140 & 0.0146\\
$\beta_{3}: 1$ & 0.8953 & 0.0193 & 0.0121 & 0.0652 & 0.0696 & 0.0363 & 0.0120 & 0.0123\\
\hline
& \multicolumn{7}{c}{ Case VI: $\varepsilon\sim N (0,1)$ with high leverage outliers}\\
$\beta_{0}: 0$ & 0.2673 & 0.2869 & 0.2901 & 0.0632 & 0.0596 & 0.0300 & 0.0114 & 0.0114\\
$\beta_{1}: 1$ & 13.2587 & 13.6355 & 13.6754 & 0.0590 & 0.0658 & 0.0305 & 0.0123 & 0.0127\\
$\beta_{2}: 1$ & 0.1817 & 0.1889 & 0.1922 & 0.0660 & 0.0727 & 0.0344 & 0.0139 & 0.0144\\
$\beta_{3}: 1$ & 0.1546 & 0.1607 & 0.1643 & 0.0668 & 0.0710 & 0.0344 & 0.0107 & 0.0108\\
\hline
\end{tabular}
\end{table}

\begin{table}[htb]
\label{tab6}\centering \caption{Cigarettes data} \vskip 0.05in
\def\arraystretch{1}
\small \hspace*{-22.75pt}
\begin{tabular}{|c|c|c|} \hline
 Country& Per capita consumption of cigarette & Deaths rates \\
 \hline
 Australia  & 480   & 180 \\
Canada  & 500   & 150 \\
Denmark & 380   & 170 \\
Finland & 1100  & 350 \\
GreatBritain    & 1100  & 460 \\
Iceland & 230   & 060 \\
Netherlands & 490   & 240 \\
Norway  & 250   & 090 \\
Sweden  & 300   & 110 \\
Switzerland & 510   & 250 \\
USA & 1300  & 200 \\
  \hline
\end{tabular}
\end{table}

\begin{table}[htb]
\label{tab7}\centering \caption{Regression estimates for Cigarettes data} \vskip 0.05in
\def\arraystretch{1}
\small \hspace*{-22.75pt}
\begin{tabular}{|c|cc|cc|} \hline
& \multicolumn{2}{c|}{Complete data}&\multicolumn{2}{c|}{Data without USA}\\
  Estimators & Intercept& Slope  & Intercept   & Slope  \\
 \hline
 LS & 67.5609& 0.2284 & 9.1393  &0.3687 \\
MM  & 7.0639 &0.3729&5.9414  & 0.3753\\
REWLSE  & 9.1393 & 0.3686& 9.1393  & 0.3686\\

  \hline
\end{tabular}
\end{table}

\begin{table}[htb]
\label{tab8} \centering\caption{Breakdown Points and Asymptotic Efficiencies of Various Regression Estimators} \vskip 0.05in
\def\arraystretch{1}
\small \hspace*{-22.75pt}
\begin{tabular}{c c c c} \hline
 & Estimator &   Breakdown Point   & Asymptotic Efficiency \\
\hline
High BP &LMS & 0.5 &0.37\\
&LTS & 0.5 &0.08\\
&S-estimates & 0.5 &0.29\\
&GS-estimates & 0.5 &0.67\\
&MM-estimates & 0.5 &0.85\\
&REWLSE & 0.5 & 1.00\\
&&\\
Low BP& GM-estimates(Mallows,Schweppe) & $1/(p+1)$ &0.95\\
&Bounded R-estimates & $<0.2$ &0.90-0.95\\
&Monotone M-estimates & $1/n$ &0.95\\
&LAD & $1/n$ & 0.64\\
&OLS & $1/n$ &1.00\\
\hline

\end{tabular}
\end{table}

\begin{figure}[h!]
    \centering
      \includegraphics[width=6in, height=5in]{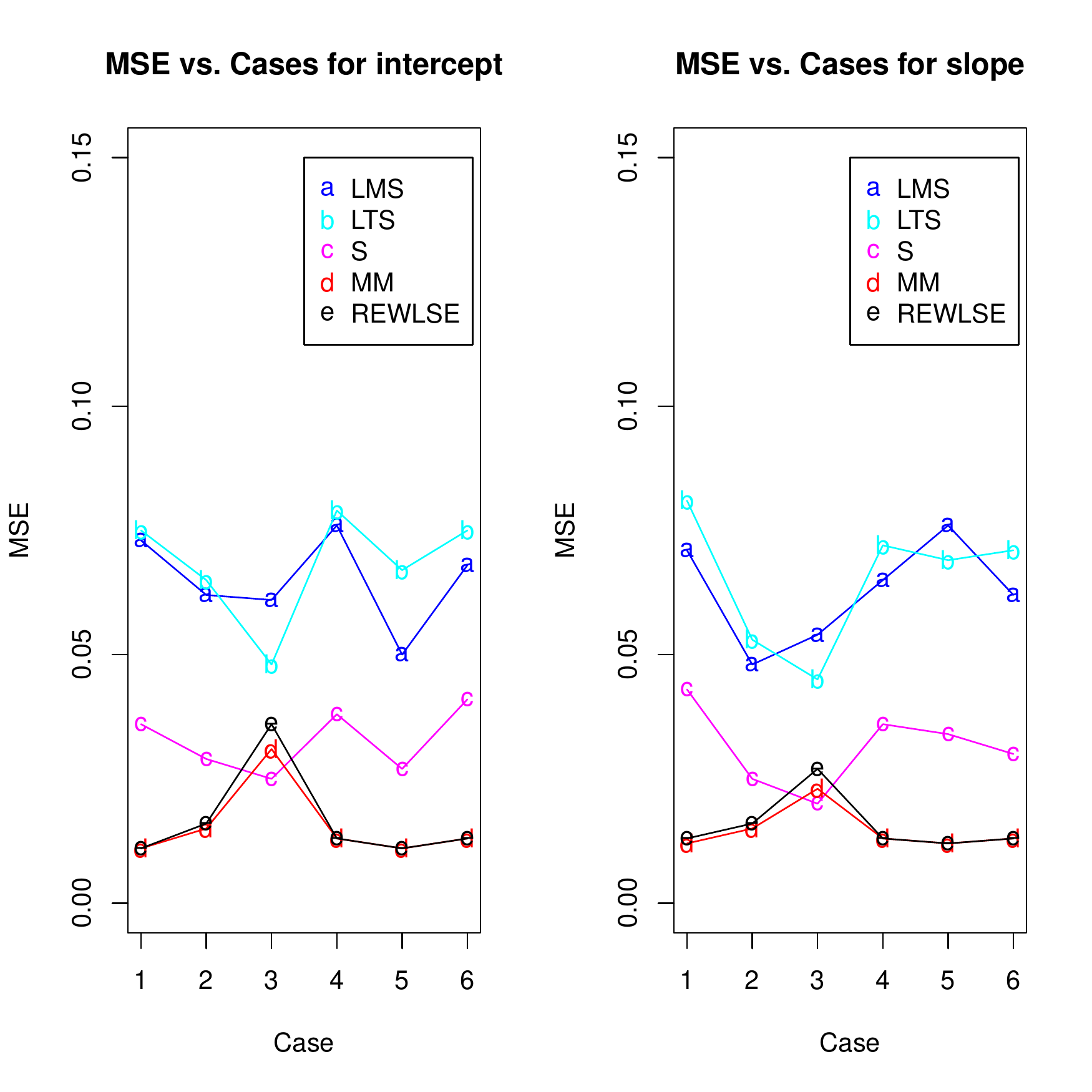}
      \caption{Plot of MSE of intercept (left) and slope (right) estimates vs. different cases for LMS, LTS, S, MM, and REWLSE, for model 1 when $n=100$.}
  \label{figure1}
\end{figure}

\begin{figure}[h!]
    \centering
      \includegraphics[width=6in, height=6in]{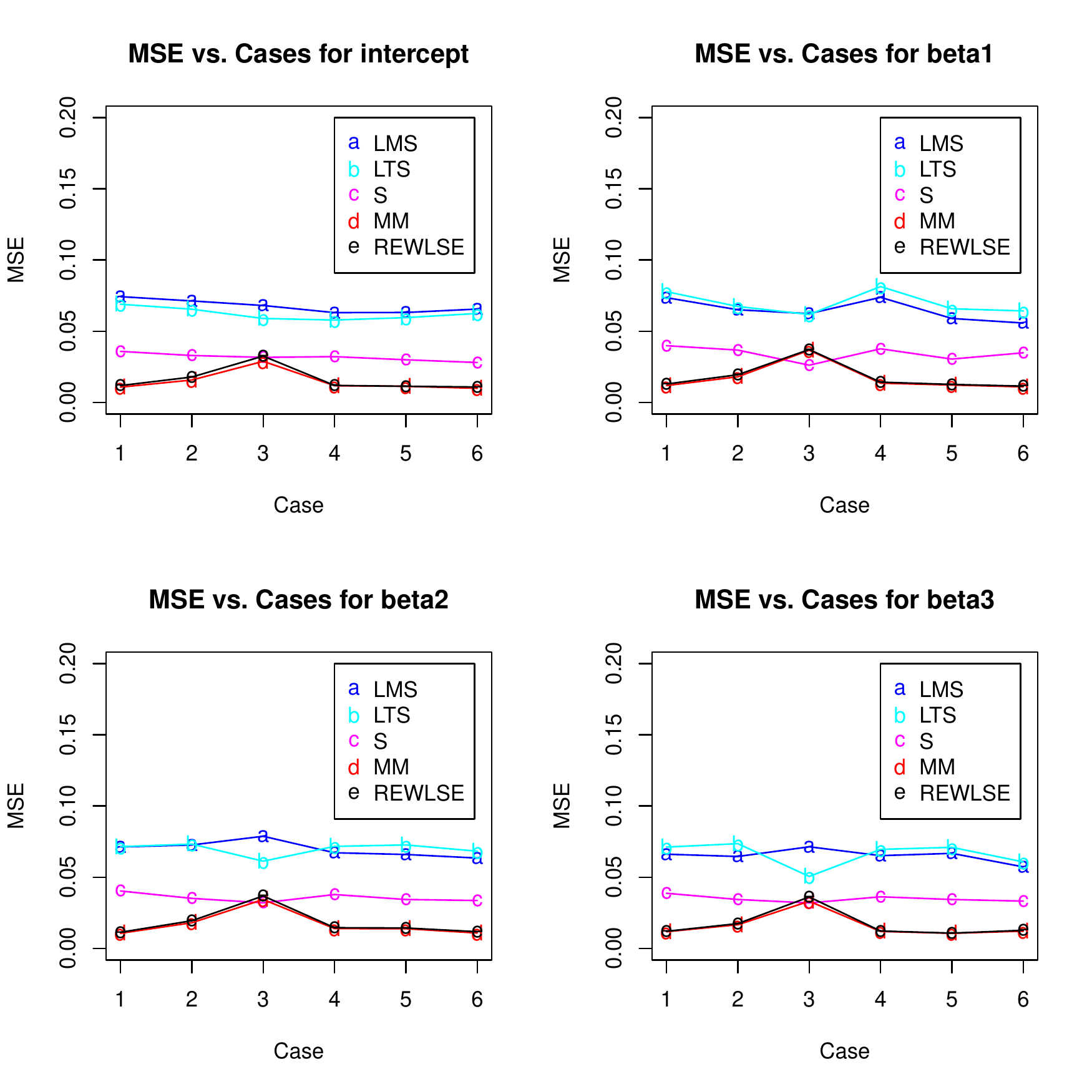}
      \caption{Plot of MSE of different regression parameter estimates vs. different cases for LMS, LTS, S, MM, and REWLSE, for model 2 when $n=100$. }
  \label{figure2}
\end{figure}

\begin{figure}[h!]
    \centering
      \includegraphics[width=6in, height=5in]{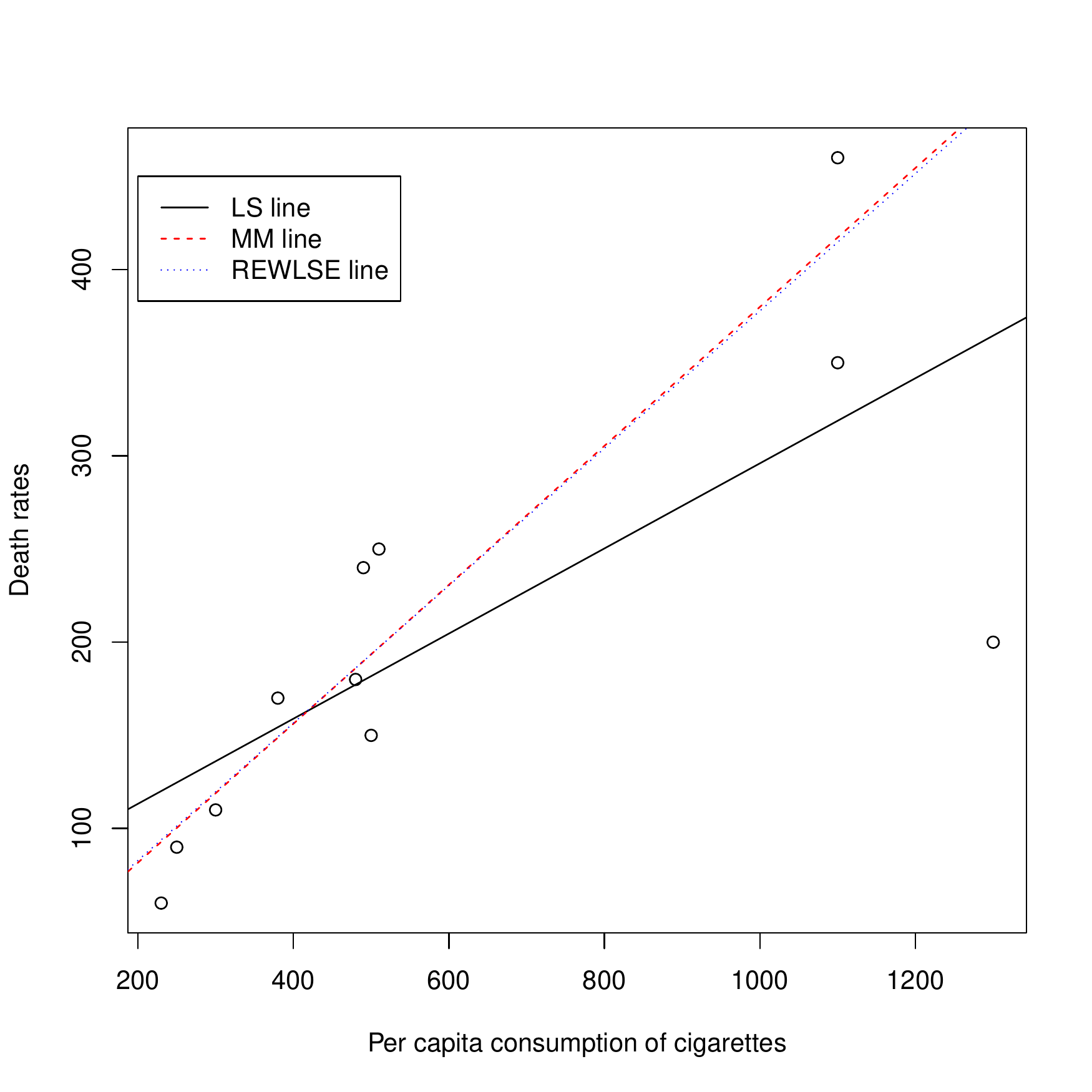}
      \caption{Fitted lines for Cigarettes data}
  \label{figure3}
\end{figure}

\end{document}